\documentclass{aa}

\usepackage{graphicx}
\usepackage{amssymb}

\begin{document}

\thesaurus{08(08.02.3;08.03.4:08.06.2:08.16.5:02.01.2)} 

\title{Accretion in Taurus PMS binaries: a spectroscopic study
  \thanks{Based on observations made with the Canada-France-Hawaii
    Telescope, operated by the National Research Council of Canada, the
    Centre National de la Recherche Scientifique de France and the
    University of Hawaii} }

\author{G. Duch\^ene\inst{1} \and J.-L. Monin\inst{1,2} \and J.
  Bouvier\inst{1} \and F. M\'enard\inst{1,3} }

\offprints{G. Duch\^ene, Gaspard.Duchene@obs.ujf-grenoble.fr}
\institute{Laboratoire d'Astrophysique, Observatoire de Grenoble,
  Universit\'e Joseph Fourier, BP 53, 38041 Grenoble Cedex 9, France \and
  Institut Universitaire de France \and Canada-France-Hawaii Telescope
  Corporation, PO Box 1597, Kamuela HI 96743, USA}

\date{Received 11 February 1999; accepted 22 September 1999}

\authorrunning{G. Duch\^ene et al.}
\titlerunning{Accretion in Taurus PMS binaries} 

\maketitle

\def\rlim{$\rho_{lim}\,$}
\def\halpha{H$\alpha\,\,$}
\def\hgamma{H$\gamma$}
\def\hdelta{H$\delta$}

\begin{abstract}
  
  We present low-resolution optical spectra of each component of 10
  T\,Tauri (TT) binary systems with separations ranging from $0\farcs9$ to
  $3\farcs5$ and located in the Taurus star-forming region. We
  derive the spectral type and \halpha equivalent width of each
  component.
  
  Complementing these results with those of Monin et al.
  (\cite{monin}) yields a sample of 14 binaries and one triple system,
  with resolved spectroscopy and/or near-infrared photometry.  We find
  that mixed binaries (CTTS+WTTS) are rare, representing only
  15--20\,\% of the systems in the separation range of $0\farcs8$ to
  $3''$. Supplementing these results with those of Hartigan et al.
  (\cite{hss}) and Prato \& Simon (\cite{prato_simon}), we show that
  the trend of binary TTS to be twins holds to separations up to
  $13''$. This is unlikely to be the result of random pairing, and
  confirms previous results that both stars in young binaries accrete
  over the same time span.
  
  In binary systems where both stars are still accreting, our
  measurements show that the most massive star is usually the
  component with the largest accretion rate by up to a factor of 10,
  as determined from the \halpha lumino\-si\-ty.

\end{abstract}


\section{Introduction}
\label{sec:intro}

During the past five years, many studies have addressed the issue of
multiplicity in low mass star-forming regions. A majority of G-K main
sequence (MS) dwarfs belong to multiple systems in the solar vicinity
(Duquennoy \& Mayor \cite{dm91}), and several studies (Leinert et al.
\cite{leinert}, Reipurth \& Zinnecker \cite{rz}, Ghez et al.
\cite{ghez}, Simon et al \cite{simon2}) have shown that this is also
the case among pre-Main Sequence (PMS) stars. The binary fraction can
vary with star formation region (SFR), and in the Taurus cloud, the
binary excess over MS stars is of the order of 1.7, indicating that
binarity is a fundamental feature of stellar formation, at least in
this SFR (see Duch\^ene, \cite{duch99}).

Amongst the various mechanisms proposed so far for binary star
formation, fragmentation appears as the most likely to meet
observational constraints (Boss \cite{boss93}). Numerical codes have
been successful in reproducing the formation of binary or multiple
systems (Bonnell et al. \cite{frag0}, Sigalotti \& Klapp
\cite{frag2}b, Boss \cite{frag3}, Burkert et al. \cite{frag1}).
However, current binary formation codes do not offer enough resolution
and time span to follow the formation and evolution of circumstellar
accretion disks. Only larger structures, which are not necessarily in
equilibrium, are predicted, providing only indirect information about
these disks, and the fate of the available circumstellar matter
remains unclear.

Various authors have studied tidal interaction of circumstellar disks
in binary systems for coplanar disks (see a review by Lin \&
Papaloizou \cite{linpapa}), and demonstrated that Lindblad resonances
create a gap in the binary environment, separating two circumstellar
disks from a circumbinary one. Accretion from the outer disk onto the
inner ones and, eventually, on both stars is prevented by
gravitational resonances.  However, Artymowicz \& Lubow (\cite{arty2})
showed that, under some hypotheses on the disk properties, matter
could flow through one or two points of the inner ring of the
circumbinary disk toward the central system. If both stars have
similar masses, both circumstellar disks are replenished, while, in
the case of very unequal masses, the accretion funnel is mainly
directed toward the secondary. On the other hand, Bonnell et al.
(\cite{frag0}) used a SPH code to study cloud fragmentation processes
and concluded that fragmentation of an elongated cloud rotating around
an arbitrary axis leads to parallel but non-coplanar accretion disk
like structures. They find that, in low mass ratio systems ($q\ll1$),
accretion of low angular momentum material is directed toward the
centre of mass, which is close to the most massive star. Thus, in
these systems, the primary appears more obscured and reddenned than
its less active companion. The different conclusions about the more
actively accreting star are likely due to the different approaches
used in these studies: while Artymowicz \& Lubow (\cite{arty2}) start
with a star+disk system to which they add a second star, Bonnell \&
Bastien (\cite{bonnell}) model the formation of such a binary from the
onset of the gravitational collapse. Also, the different initial
conditions used in these two studies imply different angular momentum
values for the accreting material (see Bate \& Bonnell
\cite{bate-bonnell}).

The study of accretion activity on both components in PMS binary
systems brings insight into the way the residual matter flows onto the
central stars. This activity can be traced by spectroscopic
measurements.  However, up to now, such studies on PMS binaries in
Taurus have been limited to wide systems (Hartigan et al. \cite{hss},
hereafter H94) due to the limited spatial resolution of the
observations. Monin et al. (\cite{monin}, hereafter paper\,I) have
started a spectroscopic survey of wide young binaries in Taurus. In
this paper, we extend this study to closer systems (down to
$0\farcs9$), investigating the classification as classical (C) or
weak-line (W) TTS of both stars in these binaries, along with a more
detailed study of the spectroscopic signature of their accretion
activity. We restrict ourself to the Tau-Aur association and we
complement our results with those of H94 and Prato \& Simon
(\cite{prato_simon}, hereafter PS97) to extend this study to a wider
range of systems.

In section\,\ref{sec:observations}, we present the observations and
the data reduction process. The results and the classification of
individual stars as C/W TTS are presented and discussed in
section\,\ref{sec:results}, and an evaluation of the random pairing
hypothesis is presented in section\,\ref{sec:pairing}.  The accretion
activity of each component within binaries is compared in
section\,\ref{subsec:otherprop}. A discussion and a summary are
presented in section\,\ref{sec:summary}.


\section{Observations} 
\label{sec:observations}

\subsection{The sample}

We have chosen our sample from the list of Mathieu (\cite{mathieu}).
In paper\,I, Monin et al. already presented some spectroscopic
measurements on five objects in this list, with separations ranging
between $2\farcs4$ and $5\farcs9$. In this paper we present
complementary observations of closer binaries from the same list. This
new sample (see Table\,\ref{tab:sample}) now includes all the binaries
in this list with separations ranging between $0\farcs89$ and
$3\farcs1$, to the exception of HBC\,411 (CoKu Tau/3) and HBC\,389
(Haro 6-10).
 
\begin{table}[t]
\caption{Complete list of spectroscopically observed binaries 
(paper\,I and this paper). Listed are the Herbig \& Bell (\cite{hbc})
catalogue numbers (hereafter: HBC) of the primary and secondary
when available, the binary separation and the previous classification
of the whole system as CTTS or WTTS (from HBC unless explicitly
quoted).}
\begin{tabular}{llll}
\hline
HBC & object & $\rho$ $(\arcsec)$ & TTS \\
\hline
356--357 & NTTS\,040012+2545\,S--N$^{\mathrm{\dagger}}$ &
1.33$^{\mathrm{a}}$ & W \\ 
358 & NTTS\,040047+2603\,W & 1.6 & W \\
379 & LkCa\,7$^{\mathrm{\dagger}}$ & 1.1 & W \\ 
& J 4872$^{\mathrm{\dagger}}$ & 3.5 & W$^{\mathrm{b}}$ \\ 
43--42 & UX\,Tau AB$^{\mathrm{\ast}}$ & 5.9 & WW \\ 
43 & UX\,Tau AC$^{\mathrm{\ast}}$ & 2.7 & W \\ 
44 & FX\,Tau$^{\mathrm{\dagger}}$ & 0.9 & C \\ 
45 & DK\,Tau$^{\mathrm{\ast}}$ & 2.8 & C \\ 
48 & HK\,Tau$^{\mathrm{\ast}}$ & 2.4 & C \\ 
& GG\,Tau/c$^{\mathrm{\dagger}}$ & 1.4 & -- \\ 
57 & GK\,Tau$^{\mathrm{\dagger}}$ & 2.5 & C \\ 
60 & HN\,Tau$^{\mathrm{\ast}}$ & 3.1 & C \\ 
& IT\,Tau$^{\mathrm{\dagger}}$ & 2.4 & C$^{\mathrm{b}}$ \\ 
73--424 & Haro\,6-37$^{\mathrm{\ast}}$ & 2.7 & CC \\ 
76 & UY\,Aur & 0.9 & C \\
80 & RW\,Aur$^{\mathrm{\dagger}}$ & 1.5 & C \\ 
\hline
\end{tabular}
\begin{list}{}{}
\item[$^{\mathrm{\dagger}}$] resolved $VRI$ imaging photometry was 
obtained for these objects
\item[$^{\mathrm{\ast}}$] paper\,I
\item[$^{\mathrm{a}}$] this work
\item[$^{\mathrm{b}}$] Hartmann et al. (\cite{h91}) \end{list}
\label{tab:sample}
\end{table}

\subsection{New spectroscopic observations} 

The observations were conducted on 1996 November 5 and 6, and December
1, at the Canada-France-Hawaii Telescope on Mauna Kea. We used the
STIS2 $2048\times2048$ detector with a $0\farcs16$/pixel scale. Using
SIS (Subarcsecond Imaging Spectrograph) providing tip-tilt correction,
we obtained an angular resolution of about $0\farcs6$ to $0\farcs8$.
Differential $VRI$ imaging photometry was also perform\-ed during the
first two nights for some targets. For each system, the primary has
been defined as the brightest star in the $V\,$band.

Long-slit spectra were obtained using a 1$\arcsec$ slit and a grism.
The usefull range of the spectra is 4000 to $7800\,$\AA, yielding a
$1.8\,$\AA\,/pixel scale. However, the actual resulting spectral
resolution is $9.6\,$\AA, except for HBC\,356--357 where it is
$12.5\,$\AA. Spectra of calibration lamps and of a spectrophotometric
standard (Feige\,110) were obtained every night. All spectra have been
wavelength calibrated, cosmic-ray cleaned, flat fielded, sky emission
subtracted and flux calibrated. All data reduction steps were
performed with standard IRAF\footnote{IRAF is distributed by the
  National Optical Astronomy Observatories, which is operated by the
  Association of Universities for Research in Astronomy, Inc., under
  contract to the National Science Foundation} routines. The two
stellar spectra of each binary were deblended and extracted
using a task fitting two gaussians with the same FWHM profile. This
reduction procedure is accurate as long as the separation remains
larger than the seeing, which was the case for all our sources except
FX\,Tau and UY\,Aur, the closest systems of our sample (see
section\,\ref{subsec:resindiv} for details).

Our estimates of the spectral types are based on the strength of TiO
bands for M stars, and on relative strengths of \ion{Ca}{i}
$\lambda\lambda$6122,62, \ion{Na}{i} $\lambda5893$, CaH
$\lambda\lambda$6350,80 and CaH $\lambda\lambda$6750--7050 for K
stars. We used the standard grids from Allen \& Strom (\cite{allen})
and Kirkpatrick et al. (\cite{kirk}), and we also observed a series of
spectral type standards during the same nights as the binary targets.
From these standard stars measurements, we find that our estimates are
accurate to within one subclass for the whole sample. However, we are
unable to determine spectral types later than M5, because most of the
spectral features we use do not change anymore with effective
temperature for such late type stars. Spectra at longer wavelengths
are needed for the classification of the reddest objects.

Uncertainties on emission line equivalent widths (hereafter EWs) were
estimated by using the maximum and minimum acceptable continuum values
next to the lines. They are typically smaller than 5\%, except for the
weakest lines, where they are of the order of 0.1--0.2\,\AA. In the
blue part of the spectrum, for the faintest stars, uncertainties can
reach 10--15\%.

We evaluated differential photometry for 8 of our sources in the $VRI$
bands. Uncertainties are usually smaller than 0.02\,mag and never
exceed 0.03\,mag.


\section{Results}
\label{sec:results}

\begin{figure*}
  \includegraphics[width=\textwidth]{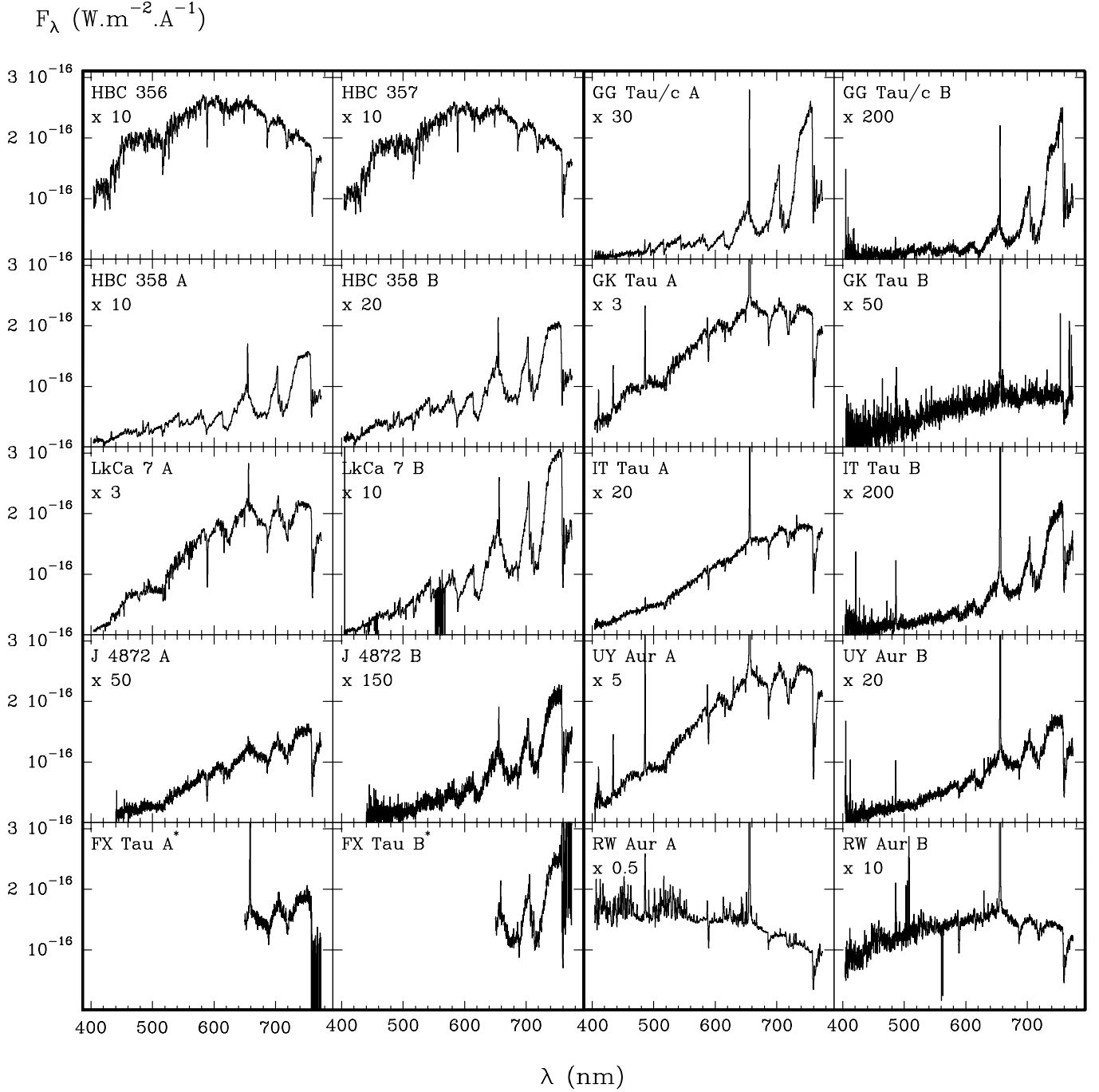}
  \caption[]{\label{fig:spectra} $F_\lambda$ spectra for all components of
    our study. The flux range is fixed for all stars, and all spectra
    have been scaled for convenience. Parts of the spectra with errors
    in gaussian fitting are not shown; exceptions are a small range
    around 5600\,\AA\, for LkCa\,7 and the reddest part of FX\,Tau
    spectra, around 7600\,\AA . Note that both FX\,Tau spectra are not
    flux calibrated, due to the deconvolution procedure.}
\end{figure*}

The spectra are shown in Figure\,\ref{fig:spectra} and the
corresponding results are summarized in Table\,\ref{tab:results}, with
the relative photometry given as $\Delta M = M_{B} - M_{A}$. For some
objects, we could also detect \hgamma, \hdelta,
[\ion{O}{i}]$\lambda6363$ and the [\ion{S}{ii}]$\lambda\lambda$6716,31
doublet in emission (see \ref{appendix}).

\subsection{Comments on individual binaries}
\label{subsec:resindiv}

The spectral type of GK\,Tau\,B could not be determined due to a poor
signal-to-noise ratio but its spectrum does not show strong TiO
absorption bands.  The spectral type of RW\,Aur\,A is undetermined
from our spectra because the star is heavily veiled by a hot continuum
and does not show any photospheric feature ; higher resolution spectra
are needed to assess its spectral type, see Basri \& Batalha
(\cite{baba}) or Chen et al. (\cite{chen}).

UY\,Aur is one of the closest binary in our sample, leading to a
possible contamination of the spectrum of the secondary by that of the
primary. We have checked this point by performing careful cuts through
the UY\,Aur spectrum perpendicular to the dispersion axis.  These cuts
show a systematic asymmetry, which position does not change with
wavelength and is not observed on the primary of any other system,
even if it is observed with the same position of the slit.
Furthermore, the separation we infer from the spectra
($0\farcs90\pm0\farcs05$) is fully consistent with the result of
near-infrared imaging (Close et al. \cite{uyao}) and the resulting
spectrum of the secondary displays different spectral features than
the primary.

In the case of FX\,Tau, the raw spectrum clearly shows two separated
peaks, but they are very close (the seeing was about $0\farcs8$ FWHM).
The double gaussian fitting procedure was unsuccessful, and we had to
apply a line by line deconvolution process. The seeing is slightly
better at longer wavelength and we could retrieve both components in
this part of the spectra only. They show significantly different
features, so we believe that we have resolved the system. This is
enough to measure the \halpha emission and to estimate the spectral
type, though with a larger uncertainty (2 subclasses).

Optical spectra of the GG\,Tau/c binary were obtained by White et al.
(\cite{white}), who found spectral types M5 and M7 for the primary and
secondary, respectively. This is in agreement with our findings for
both components, although we could not determine accurately the
spectral type of the secondary.

\begin{table*}
\caption{Photometric and spectroscopic results for stars in our 
sample. ``--'' means that the line is undetected in our spectra. 
The relative photometry given in parentheses was obtained from our 
spectra (see section\,\ref{subsec:otherprop}). No emission line has 
been measured in the spectrum of FX\,Tau at wavelengths shorter than
 6500\,\AA. The last column present our classification of the 
stars: ``W'' for WTTS and ``C'' for CTTS (see text for details).} 
\begin{tabular}{llccccccccc}
\hline
HBC & object & spec. type & $\Delta V$ & $\Delta R$ & $\Delta I$ & \halpha 
& H$\beta$ & [\ion{O}{i}]$\lambda6300$ & \ion{He}{i} $\lambda6678$ & type \\
& & & (mag) & (mag) & (mag) & EW$_\lambda$(\AA) & EW$_\lambda$(\AA) &
EW$_\lambda$(\AA) & EW$_\lambda$(\AA) & \\ 
\hline
356 & NTTS\,040012+2545\,S & K3 & 0.07 & 0.08 & 0.06 & 0.4 
& -- & -- & -- & W \\
\vspace*{0.1cm}
357 & NTTS\,040012+2545\,N & K3 & (0.1) & (0.0) & & 0.4 & -- & -- & -- & W \\
358 & NTTS\,040047+2603\,W\,A & M5 & & & & 10.0 & 8.7 & -- & -- & W \\ 
\vspace*{0.1cm}
& NTTS\,040047+2603\,W\,B & M5 & (0.3) & (0.3) & & 6.7 & 6.0 & -- & -- & W \\
379 & LkCa\,7\,A & K9 & 2.18 & 2.03 & 1.03 & 2.5 & -- & -- & -- & W \\ 
\vspace*{0.1cm}
& LkCa\,7\,B & M4 & (2.0) & (1.8) & & 5.6 & -- & -- & -- & W \\
& J\,4872\,A & K9 & 1.50 & 1.40 & 0.83 & 0.8 & -- & -- & -- & W \\ 
\vspace*{0.1cm}
& J\,4872\,B & M1 & (1.7) & (1.5) & & 4.2 & -- & -- & -- & W \\
44 & FX\,Tau\,A & M1 & 0.24 & 0.30 & $-$0.10 & 13 & & & -- & C \\ 
\vspace*{0.1cm}
& FX\,Tau\,B & M4 & & & & 1.0 & & & -- & W \\ 
& GG\,Tau/c\,A & M5 & 2.74 & 2.57 & 2.13 & 22 & 15 & -- & -- & C \\ 
\vspace*{0.1cm}
& GG\,Tau/c\,B & $>$M5 & (2.6) & (2.5) & & 19 & 39 & -- & -- & C \\ 
57 & GK\,Tau\,A & K7 & 4.26 & 4.20 & 4.10 & 35 & 13 & -- & 0.3 & C \\
 \vspace*{0.1cm}
& GK\,Tau\,B & -- & (4.1) & (4.1) & & 45 & 11 & -- & -- & C \\
& IT\,Tau\,A & K3 & 3.63 & 3.15 & 2.27 & 21.7 & 3.4 & -- & -- & C \\ 
\vspace*{0.1cm}
& IT\,Tau\,B & M4 & (3.6) & (3.2) & & 147. & 52. & 0.3 & 0.7 & C \\
76 & UY\,Aur\,A & K7 & & & & 86. & 29. & 1.5 & 0.8 & C \\
\vspace*{0.1cm}
& UY\,Aur\,B & M2 & (2.7) & (2.5) & & 69.1 & 33.8 & 3.8 & -- & C \\ 
80 & RW\,Aur\,A & ?e & 4.36 & 3.85 & 3.25 & 76 & 7.9 & 1.1 & 1.3 & C \\ 
& RW\,Aur\,B & K5 & (3.4) & (3.2) & & 42.7 & 7.3 & 2.1 & 0.3 & C \\ 
\hline
\multicolumn{10}{c}{Data from paper\,I }\\
\hline
43 & UX\,Tau\,A & K4 & & & & 9.5 & -- & -- & -- & C \\ 
42 & UX\,Tau\,B & M2 & (2.4) & (2.2) & & 4.5 & 3.5 & -- & -- & W \\ 
\vspace*{0.1cm}
& UX\,Tau\,C & M3 & (4.0) & (3.9) & & 8.5 & -- & -- & -- & W \\ 
45 & DK\,Tau\,A & K9 & & & & 31. & 16. & 2.5 & 0.7 & C \\ \vspace*{0.1cm}
& DK\,Tau\,B & M1 & (1.7) & (1.6) & & 118. & 33.5 & 4.8 & 1.5 & C \\ 
48 & HK\,Tau\,A & M1 & & & & 50. & 20. & -- & -- & C \\ \vspace*{0.1cm}
& HK\,Tau\,B & M2 & (2.6) & (2.7) & & 12.5 & -- & -- & -- & C \\ 
60 & HN\,Tau\,A & ?e & & & & 230. & 64. & 16. & 2.5 & C \\ \vspace*{0.1cm}
& HN\,Tau\,B & M4.5 & (3.8) & (3.7) & & 65. & 41. & 8. & 2 & C \\ 
73 & Haro\,6-37\,A & K8 & & & & 19.5 & 3.7 & -- & -- & C \\ \vspace*{0.1cm}
424 & Haro\,6-37\,B & M0 & (1.4) & (1.2) & & 195. & 60. & -- & -- & C \\ 
\hline
\end{tabular}
\label{tab:results}
\end{table*}

We have also determined an accurate estimate of the separation of
HBC\,356--357: $1\farcs33 \pm 0\farcs05$. Walter et al.
(\cite{walter}) reported a somewhat larger separation (2$\arcsec$).
However, these authors did not publish the uncertainty on their
result, and we believe that this discrepancy is unlikely to be due to
orbital motion.

In order to study the relative accretion activity of the individual
components of the binaries of our sample, we first determined which
stars actually accrete, i.e. the respective classification of the
observed stars as CTTS or WTTS. In the following, we use every
available piece of information to establish this classification.

\subsection{TTS classification criteria} 
\label{subsec:classification}

The first large scale surveys for TTS were objective prism surveys and
the ``historical'' criterion to detect a CTTS used the \halpha EW by
checking whether it was larger than 10\,\AA\, or not (e.g. Strom et
al. \cite{S89}). The stars identified as TTS from their photometry,
but with smaller \halpha EWs were classified as WTTS, i.e. non-active
PMS stars. However, this threshold is not a sharp edge, and a more
physically meaningful diagnosis would be the \halpha flux (Cohen \&
Kuhi, \cite{ck}). Moreover, Mart\'{\i}n (\cite{martin98}) discussed
the possibility that the \halpha EW threshold varies with spectral
type, later spectral types stars having a higher threshold. He
proposed a 5\,\AA\, EW limit for K stars and 10\,\AA\, for early M
stars. We adopt this criterion in our classification.

We have also checked this classification against other criteria such
as [\ion{O}{i}]$\lambda6300$ emission line and $K-L$ infrared excess.
Edwards et al. (\cite{edwards}) have found that all stars with
detectable [\ion{O}{i}] emission or $K-L$ excess ($>0.4$)
systematically have \halpha EWs larger than 10\,\AA.

However, in order to compare our newly classified TTS with previously
known field TTS, the use of different criteria may lead to confusion
and unexpected biases. This point will be carefully examined below
(see section\,\ref{pairing:bias?}), but we stress that only one star
out of the 31 listed in Table\,\ref{tab:results} has a discrepant
classification when using different criteria (see also
section\,\ref{subsubsec:diff-crit}).

\subsection{Classification of individual stars}
\label{subsec:classindiv}

\noindent {\bf NTTS\,040012+2545, 040047+2603, LkCa\,7 and
  J\,4872:} no component in any of these systems shows evidence of
accretion activity, as they all exhibit only low \halpha emission and
no other emission line, to the exception of HBC\,358. These four
systems are thus considered as WW binaries.

\noindent{\bf GK\,Tau, IT\,Tau, UY\,Aur, and RW\,Aur:} all of these
systems contain stars with moderate to strong \halpha and H$\beta$
emission and, for some of the stars, metallic and forbidden lines.
All the stars in these systems can thus be safely classifed as CTTS.

\noindent{\bf FX\,Tau: } the secondary shows a very low \halpha
emission and is probably a WTTS. On the other hand, the primary shows
a moderate emission in this line, as well as H$\beta$ emission (Cohen
\& Kuhi \cite{ck}). Furthermore, Strom et al. (\cite{S89}) and
Skrutskie et al. (\cite{S90}) reported moderate $\Delta K$ and $\Delta
N$ excesses for the system. All these evidences support the idea that
the primary is a CTTS.

\noindent{\bf UX\,Tau and HK\,Tau: } UX\,Tau\,A was observed in
paper\,I and classified as a WTTS from its \halpha EW of $9.5\,$\AA.
In such a star apparently ``at the border'' between C and WTTS, we
reexamined this classification and propose to (re)classify it as a
CTTS, because of the large \halpha emission flux (for a spectral type
K4, the CTTS threshold is only about $5\,$\AA). Another clue
is its significant $\Delta N$ excess (Skrutskie et al. \cite{S90}).
This post facto reconsideration of classification criteria can be
misleading at first sight, but we stress that it is only done here to
define a more accurate picture of an accreting TTauri star. The
classification of all other stars is identical to paper\,I. In
particular, our previous classification of HK\,Tau\,B as a CTTS
(\halpha EW of 12.5\,\AA) has been recently confirmed by an HST image
of this star showing a remarkable edge-on circumstellar disk
(Stappelfeldt et al. \cite{stap98}).


\section{CTTS - WTTS pairing within Taurus binaries} 
\label{sec:pairing}

In the following, we call ``twins'' the systems where the TTS are of
the same type (either CC or WW), and ``mixed'' the systems where the
stars are different.

One of our objects contains stars physically associated in a multiple
system (UX\,Tau). Although this system is not strongly hierarchical
($5\farcs9$ - $2\farcs7$), we consider that it can be split into two
``independent'' binaries, leading to a total of 16 binaries in our
sample. The validity of this assumption will be evaluated in
Section\,\ref{subsec:bias-comp}.

\subsection{Testing the random pairing hypothesis} 
\label{pairing:results}

The 16 binaries considered here can be divived into three categories:
9 binaries contain only CTTS ($\approx 56\,$\%), 4 are formed of two
WTTS (25\,\%), and 3 are mixed systems, all with a CTTS primary and
two of them in the same triple system, UX\,Tau, representing less than
19\,\% of our sample. Mixed systems appear to be rare in TTS binaries,
and this is even more striking when we use the ``historical'' \halpha
$10\,$\AA\, EW criterion. Then only one mixed system remains among 16
binaries, and the proportion drops to about 6\%. We use this sample to
address the question: are binary components taken at random from the
TTS population?

If we want to compare this result with a distribution randomly taken
from a single TTS population, we need to know the ratio of
WTTS-to-CTTS in Taurus. In a study limited to the central parts of the
Tau-Aur dark cloud, Hartmann et al. (\cite{h91}) found a ratio close
to unity. Considering a larger sky area leads to an even larger
WTTS-to-CTTS ratio, mainly because of the widespread $ROSAT$
population (e.g. Wichmann et al. \cite{wichmann}). Since our sample
mostly contains systems in the center of the molecular cloud, we
conservatively adopt a W/C ratio $=1$.

Taking a fixed distribution of primaries (4 WTTS and 12 CTTS), the
probability to get 3 mixed systems out of 16 binaries from randomly
taken secondaries is $\rm C^3_{16}(\frac{1}{2})^{16}\!\le 1\,\%$. We
therefore reject the hypothesis that components of TTS binaries are
randomly associated from the distribution of single stars. In other
words, the TTS types of Taurus binary components are significantly
correlated.

\subsection{Possible sources of bias}
\label{pairing:bias?}

In this section, we discuss some possible sources of errors in our
result on a preferential CC pairing in TTS binaries.

\subsubsection{The use of different classification criteria}
\label{subsubsec:diff-crit}

In section\,\ref{subsec:classification}, we have used complementary
criteria to establish the C/W TTS nature of our sources.  However,
considering only the ``historical'' $10\,$\AA\, \halpha EW
classification criterion does not only imply minor changes (1 star in
UX\,Tau is modified upon 31), but makes mixed binaries even more rare:
only one mixed pair (FX\,Tau) out of 16 remains. The probability that
the observed C/W distribution in our 16 binaries results from random
pairing then falls to $\approx 0.02\,\%$.

\subsubsection{The case of WW pairs}

The evolutionnary status of the WTTS population identified from the
$ROSAT$ All-Sky Survey is somewhat uncertain: some of these stars may
be unrelated to the TTS population (e.g. Favata et al.
\cite{favata}). If they are young main sequence stars, we expect that
both components will mimic WTTS since they are too old to still be
accreting.  Then the observation of such binaries can lead to a bias
towards WW pairs in our study. This can potentially affect 2 binaries
in our sample, which were first detected by $ROSAT$ (their names
starts with ``NTTS''). If we exclude {\it all\/} WW binaries for
safety, we end with at most 3 mixed systems out of 12, yielding a
proportion of 25\,\% mixed systems in TTS binaries. Then the
probability that this distribution results from random associations is
only $\approx 5\,\%$.  We therefore conclude that the high proportion
of twin binaries in our sample is not strongly affected by the
presence of spurious WW binaries.

\subsubsection{Time evolution}

Since the proportion of stars surrounded by a circumstellar disk
decreases with age, we inspect the possibility that our binary
population is younger, on average, than the population of singles. In
such a case, we would expect to find more CTTS (``young and active'')
than in the singles sample and, consequently, more CC binaries. In
Simon \& Prato's (\cite{simon_prato}) study, the median age of their
single stars sample is $\log t_{\rm SW}\!\sim\!5.8$. In our sample, we
find that half of the primaries that have an age assigned by Simon \&
Prato are older than this value. We thus conclude that our study
includes about as many young systems as old systems, and time
evolution effects do not impinge our conclusion.

\subsubsection{Close companions and hierarchical systems}
\label{subsec:bias-comp}

The issue of how to treat known binaries which we do not resolve is
not straightforward. Moreover, currently undetected companions may
exist around some of the stars in our sample. These unresolved
companions may strongly impact on the evolution and the accretion
history of their associated star. Furthermore, considering a triple
system as two independent binaries may not be a valid hypothesis.

To evaluate the impact of such multiple systems, we considered a
subsample of binaries where no third component, either spectroscopic
(Mathieu \cite{mathieu}), very tight visual (Simon et al.
\cite{simon2}) or wider, is known so far. To our knowledge, only 7
binaries in our overall sample match this criterion: LkCa\,7, FX\,Tau,
DK\,Tau, HK\,Tau, IT\,Tau, HN\,Tau and UY\,Aur. This subsample
contains 6 twins and 1 mixed binaries. Once again, only about $15\,\%$
of these binaries are mixed, leading us to think that the possible
existence of additional companions does not significantly modify the
results.

\subsection{Complementary results from the literature}

We have considered previous results in the literature providing
information on the classification of the components of more PMS
binaries in the Taurus SFR. We complement our results with those of
H94 and PS97 and obtain a sample that contains over 90\,\% of all known
binaries located in Taurus in the separation range
$0\farcs8$--$13\arcsec$. The list of these supplementary objects is
given in Table\,\ref{tab:wide}.

\begin{table}
\caption{Binaries observed by H94 and PS97, 
listed with the primary first. See text for details about the
classification of individual stars}
\begin{tabular}{lllcl}
\hline
HBC     & Object        & $\rho (\arcsec)$ & type & ref \\
\hline
352--353 & NTTS\,035120+3154\,SW--NE &  8.7     & WW & H94 \\
355--354 & NTTS\,035135+2528\,SE--NW &  6.3     & WW & H94 \\
360--361 & NTTS\,040142+2150\,SW--NE &  7.2     & WW & H94 \\
386--387 & FV\,Tau--FV\,Tau/c        &  12.3    & CC & H94 \\
389      & Haro\,6-10                &  1.2     & CC & PS97 \\
51--395  & V710\,Tau\,N--S           &  3.2     & CW & H94 \\
52--53   & UZ Tau\,E--W              &  3.8     & CC & H94 \\
54       & GG\,Tau--GG,Tau/c         &  10.3    & CC & H94 \\
56--57   & GI--GK\,Tau               &  12.2    & CC & H94 \\
411      & CoKu\,Tau/3               &  2.0     & CC & PS97 \\
\hline
\end{tabular}
\label{tab:wide}
\end{table}

To classify the members of binaries studied by H94, we used their
\halpha EWs and spectral types together with the estimated
near-infrared excesses. Both indicators agree well for all stars
except for V710\,Tau\,S. This star presents an \halpha EW hardly above
the classical limit (11\,\AA), with a spectral type M3, and no
infrared excess. Moreover, Cohen \& Kuhi (\cite{ck}), measured an
\halpha EW of 3.3\,\AA\, and no emission in the forbidden lines,
leading us to classify this star as a WTTS. V710\,Tau consequently
happens to be one of the few mixed pairs (CW) among TTS binaries.

For the two binaries studied by PS97 and included in our sample, all
stars have $K-L\geq1.2$\,mag. Such high values are strong evidences
for the presence of an optically thick accretion disk in the inner
0.5\,AU around each star (the upper limit for photospheric colors is
$K-L\approx0.4$\,mag, Edwards et al. \cite{edwards}), so that these
stars can be safely classified as CTTS. It is also worth mentionning
that for all systems common to the PS97's sample and ours (DK\,Tau and
UY\,Aur with $KL$ photometry and Haro\,6-37 with Br$\gamma$
spectroscopy), their classification and ours are fully consistent.

If we take these complementary results into account, we obtain a
sample of 26 binaries with 15 CC twins, 7 WW twins, and 4 mixed. The
proportion of mixed systems is then $\approx 15\%$, and even only
$\approx 4\,\%$ if we adopt the \halpha EW criterion, yielding similar
results as in Section\,\ref{pairing:results}.


\section{Differential accretion in CC binaries} 
\label{subsec:otherprop}

For all binaries where both stars have active accretion disks (CC
pairs), we have used the available spectra to compare the accretion
activity of each component, using their \halpha flux as an accretion
diagnosis. The \halpha EW has already been shown to correlate well
with the infrared excess in CTTS (eg. Edwards et al., \cite{edwards}).
Moreover, recent studies in the near infrared where the extinction is
about ten times smaller than at \halpha wavelengths, have revealed
tight correlations between the accretion luminosity and the Pa$\beta$
and Br$\gamma$ emission fluxes (Mu\-ze\-rol\-le et al.,
\cite{muzerolle}). We will then hereafter use the ratio of \halpha
fluxes in binaries, assuming that this flux is proportional to the
energy dissipated in the accretion process, i.e. to the accretion
luminosity.

We also assume that the extinction toward both components of the
binary is the same, based on the tight correlation observed in the
data of H94 between $A_J$ toward the primary and the secondary. We
checked that this correlation is still valid at smaller separations:
we evaluated a rough $V\,R$ photometry from our spectra and compared
the results to dwarfs colors. Due to observational uncertainties
($\sigma_V\!\approx\!\sigma_R\!\approx\!0.1\,$mag and 1 subclass for
the spectral type), the final accuracy of the extinction is rather
poor (typically, $\sigma(A_V)\!\approx\!0.7\,$mag). However, we did
not find any evidence that the correlation is modified. This
correlation is likely due to the fact that both components of a binary
system are equally embedded in the Taurus molecular cloud, but other
explanations include the existence of a common circumbinary envelope
and/or of circumstellar disks with similar orientations. Brandner \&
Zinnecker (\cite{bz}) reported a similar correlation for close
($<250\,$AU) PMS binaries in southern SFRs.

\begin{figure}
\includegraphics[angle=270,width=\columnwidth,origin=br]{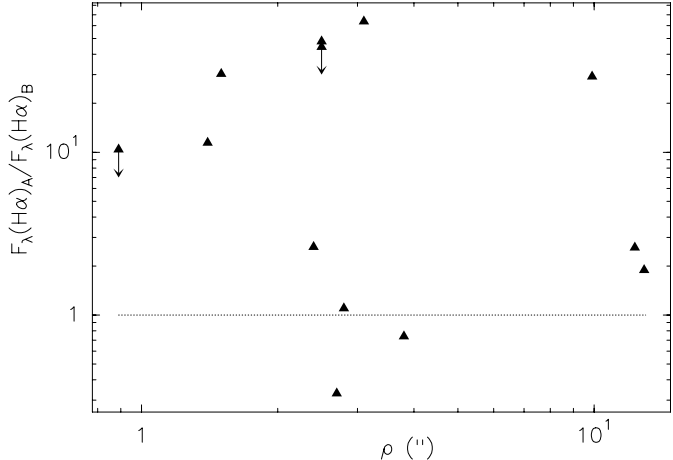} 
\caption{\halpha luminosity ratio for all CC binaries. Upper limits 
are for UY Aur and HK Tau.}
\label{fig:sep_halpha}
\end{figure}

For each binary, the ratio of the \halpha luminosities is computed as
follows:
$$\frac{F_{H_\alpha}^A}{F_{H_\alpha}^B} =
\frac{EW_{H_\alpha}^A}{EW_{H_\alpha}^B} \times
\frac{F_c^A}{F_c^B}10^{0.4(A_V^A-A_V^B)} \approx 
\frac{EW_{H_\alpha}^A}{EW_{H_\alpha}^B} \times \frac{F_c^A}{F_c^B}$$
where $F_c$ is the nearby continuum flux estimated from our spectra
when available and from H94's $R$ photometry otherwise.

Figure\,\ref{fig:sep_halpha} shows a clear trend for the primaries to
have a higher \halpha flux than the secondaries thus a higher
accretion luminosity. It is unlikely that this result is the
consequence of a systematic bias introduced by the assumption that
both extinctions are the same. This would imply that we systematically
overestimated the \halpha luminosity ratios by a factor of 4, which
requires that the extinctions toward the secondaries are larger by
about $A_V^B - A_V^A\sim$1.5\,mag. Such a systematic trend would have
been detected in H94 (the authors quote $\sigma(\Delta
A_V)\approx0.3$ mag), as well as in our data. This suggests that the
accretion rate is larger on the more massive component of the system.

In a few cases, clues exist that the photosphere is not seen directly,
and that we only detect scattered light. This is the case for UY\,Aur
B (Close et al. \cite{uyao}) and HK\,Tau B (Stapelfeldt et al.
\cite{stap98}) where an edge-on disk has been recently detected. For
these stars, the observed photometry is therefore only a lower limit
to their actual flux, and we consequently underestimated their \halpha
flux. Arrows have been accordingly added in
Fig.\,\ref{fig:sep_halpha}. No nebular structure is seen at high
angular resolution around any other star so far and we assume that
there is no object with strong scattering in our sample appart from
these two stars.


\section{Discussion and summary}
\label{sec:summary}

We have shown that there exists only few mixed systems in Taurus PMS
binaries in the separation range 130 - 1800 AU. This result extends
those of PS97 who did not find any mixed system in a sample including
binaries with separations of 40--360\,AU. This indicates that the
accretion history of the two stars are not independent, even for
binaries with separations up to $800\,$AU (from our new spectroscopic
observations) and even $1800\,$AU if we take into account the results
from H94.

What can explain such a correlation in binaries with separations as
large as $1800\,$AU ?  This ``twinning'' trend, together with the fact
that circumstellar disk dissipation times from optically thick to
optically thin are short (Simon \& Prato \cite{simon_prato}), led PS97
to propose that both components of a close binary system accrete over
the same time span because their circumstellar disks are replenished
by material from a common (circumbinary) environment.  As soon as this
environment is cleared, both disks dissapear over a short viscous
timescale.  However, the circumbinary environment hypothesis appears
difficult to apply to wide binaries, and if such envelopes have been
detected around a few close binaries, they generally remain elusive.
Similarly, it appears unprobable that the binary as a whole can sweep
enough material during its wander through the parent cloud: at
1\,km.s$^{-1}$, a 100\,AU radius wide binary sweeps only
$10^{-12}\,M_\odot.yr^{-1}$ in a $10^2\,$cm$^{-3}$ density cloud.

On the other hand, we find that the primaries have larger \halpha
fluxes than their secondaries. We call 'primary' the brightest
component in the $V$ band, which always has an earlier spectral type
than the secondary so that it is likely the most massive star.  The
\halpha luminosity is assumed to be proportional to the accretion
luminosity:
$$L_{\rm H\alpha}\propto L_{\rm a\rm c\rm
  c}=\frac{GM_\star\dot{M}}{R_\star}$$

Baraffe et al.'s (\cite{isa}) evolutionary models show that two
2\,Myr-old TTS with masses of 1\,$M_\odot$ and 0.1\,$M_\odot$ have
$M_\star /R_\star$ ratios only differing by a factor of 4 (the most
massive star also has a larger radius). Our measured \halpha
luminosity ratios vary by over 2 orders of magnitude and therefore
cannot be accounted for by extreme mass ratios. The difference in the
accretion luminosities is thus likely to reveal that, in most cases,
the primary accretes more than its companion:
$\dot{M}_A\!>\!\dot{M}_B$. It is also noticeable that the mixed
systems in our sample all have a CTTS primary, so that in the case of
CW pairs, the more massive star again seems to be more active than the
other one.

If both components have similar circumstellar disk lifetimes ($\tau_D =
M_D/\dot{M}$), these results suggest that the circumprimary disk is
preferentially feeded in the early binary formation process by a
common circumbinary reservoir of mass. This is in agreement with the
prediction of Bonnell et al.'s (\cite{frag0}) model.

Another possibility is that the accretion rate on the star, $\dot{M}$,
is proportional to the disk mass, itself related to the mass of the
central star. In the canonical accretion disk theory, the accretion
rate is related to the surface density $\Sigma$, itself evidently
linked to the disk mass. This mechanism would explain simultaneously
why $\dot{M}_A\!>\!\dot{M}_B$ and why the disk lifetime $\tau =
M/\dot{M}$ does not depend on the mass of the central star.  If true,
such a $M - \dot{M}$ relation should hold for single TTS but current
mass determination lack the precision needed to study this point
further.

Observations of closer binaries down to separations of the order of
the peak value in the PMS separation distribution ($\approx 50$\,AU,
Mathieu \cite{mathieu}) should shed more light on this question. Such
observations are within reach of current adaptive optics systems
equiped with spectroscopic capabilities. Such a peak separation is of
the order of the size of a canonical accretion disk and these
observations would allow to study systems where the star-disk and
disk-disk interactions are strong, and also where the eventual
leftover circumbinary environment has a major influence.

\begin{acknowledgements}
  We are grateful to Caroline Terquem and Mike Simon for enriching
  discussions. Comments from an anonymous referee helped to
  significantly improve this paper. This research has made use of the
  Simbad database, operated at CDS, Strasbourg, France, and of the
  NASA's Astrophysics Data System Abstract Service.
\end{acknowledgements}


\appendix

\section{Complementary line measurements } 
\label{appendix}

In our new spectroscopic observations, some emission lines which are
not presented in Table\,\ref{tab:results} were detected in some of our
targets. EWs measurements for H$\gamma$, H$\delta$,
[\ion{O}{i}]$\lambda6363$ and the doublet
[\ion{S}{ii}]$\lambda\lambda6716,31$ lines are given in
Table\,\ref{tab:append} for the stars where these lines were detected.

\begin{table}
\caption{Additionnal emission line measurements from our spectra. 
All equivalent widths are given in A. Stars that are not listed 
here were not detected in any of these lines.}
\begin{tabular}{llcccc}
\hline
HBC & object & H$\gamma$ & H$\delta$ & [\ion{O}{I}]
& [\ion{S}{II}] \\
    &        &           &           & $\lambda$6363    
& $\lambda\lambda$6716,31 \\
\hline
358 & HBC\,358\,A & 9.5 & 5.1 & -- & -- \\
    & HBC\,358\,B & 4.8 & 11.5 & -- & -- \\
    & GG\,Tau/c\,A & 12 & -- & -- & -- \\
57  & GK\,Tau\,A & 10.2 & 7.0  & -- & -- \\ 
    & IT\,Tau\,A & 3.4 & -- & -- & -- \\
    & IT\,Tau\,B & 35$\pm$5 & -- & -- & -- \\ 
76  & UY\,Aur\,A & 24 & 8.0$\pm$1.0 & -- & -- \\ 
    & UY\,Aur\,B & 13 & -- & 2.1 & 1.2 \\ 
80  & RW\,Aur\,A & 1.1 & 3.1 & -- & -- \\
    & RW\,Aur\,B & -- & -- & 0.8 & -- \\
\hline
\end{tabular}
\label{tab:append}
\end{table}


\end{document}